\documentclass[11pt,twoside]{article}
\usepackage{cozumel2005}
\usepackage{graphicx}
\def\ri{\hbox{$R\!-\!I$}}
\def\vk{\hbox{$V\!-\!K$}}
\def\bvi{\hbox{$BV\!I$}}
\def\bvrijhk{\hbox{$BV\!RI\!J\!H\!K$}}
\def\bvri{\hbox{$BV\!RI$}}
\def\jhk{\hbox{$J\!H\!K$}}

\begin{document}
\title{Cepheids, Eclipsing Binaries and the Distance Scale: from the Galaxy to the Local Group}    
\author{Lucas M. Macri\altaffilmark{1}}   
\affil{National Optical Astronomy Observatory, 950 North Cherry Avenue, Tucson, AZ 85719, United States}    
\altaffiltext{1}{Hubble Fellow}
\begin{abstract} 
I present a review of recent observational work on Cepheid variables and eclipsing binaries in our Galaxy, the Magellanic Clouds, and other members of the Local Group.
\end{abstract}


\section{Introduction}                      
The SOC has kindly invited me to review recent work on Cepheid variables and eclipsing binaries, and their impact on the Extragalactic Distance Scale. They have asked me to focus on our Galaxy, the Magellanic Clouds, and other members of the Local Group. Given the large amount of ongoing work on these topics, this review may inadvertently have missed a recent article or two, or failed to mention a seminal contribution from earlier times. I apologize in advance for any such omissions, and urge the interested reader to follow up on the articles listed in bibliography for additional references to work on the field. I also recommend the review by \citet{jacoby92} on the Extragalactic Distance Scale as well as the review by \citet{andersen91} on eclipsing binaries, both of which contain extensive references to previous work on the field. 

This article is organized as follows: \S 2 discusses advances on Galactic Cepheids; \S 3 summarizes work on Magellanic Cloud Cepheids and detached eclipsing binaries; \S 4 reviews discoveries in other Local Group members. \S 5 briefly examines research focused on more distant galaxies and \S 6 contains the conclusions.

\section{Cepheids in the Milky Way Galaxy}

\subsection{Parallaxes}

The {\it Hipparcos} satellite \citep{perryman97} provided parallaxes for over 200 Cepheid variables, but most of the measurements had extremely low $S/N$. \citet{feast97} carried out the first determination of the Cepheid Period-Luminosity relation zeropoint based on {\it Hipparcos} parallaxes, followed by \citet{madore98}, \citet{lanoix99} and \citet{groenewegen00}. 

Recently, Benedict and collaborators have carried out a campaign to obtain high-precision parallaxes for 11 nearby Cepheids using the Fine Guidance Sensors onboard the {\it Hubble Space Telescope}. Their first result was a new parallax for $\delta$~Cep \citep{benedict02}. Observations for an additional 10 Cepheids have been obtained and are currently being analyzed. The {\it HST} data will enable a direct calibration of the Galactic Cepheid P-L relation.

\begin{figure}
{\centerline{\includegraphics[width=4.6in]{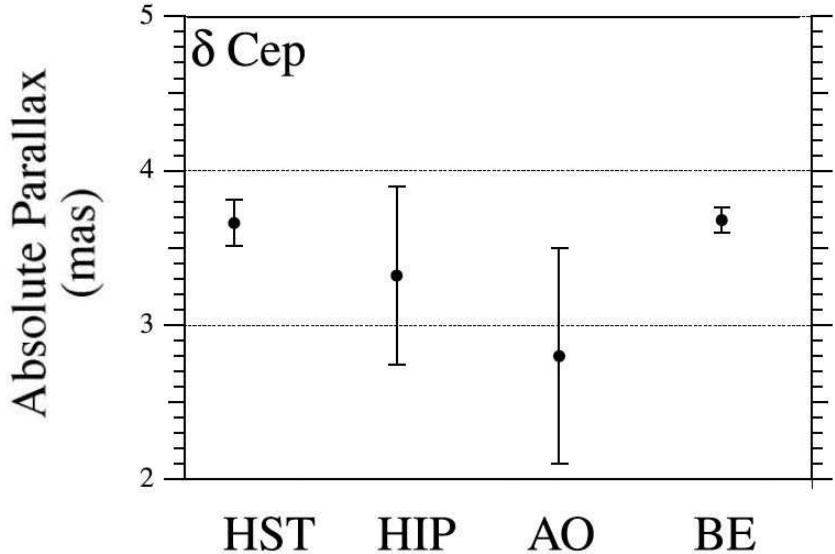}}}
\caption{Reproduction of Figure 6 from \citet{benedict02}. Recent parallax measurements of $\delta$ Cephei. HST=their results; HIP={\it Hipparcos} measurement; AO=\citet{gatewood98}; BE=predicted parallax from Barnes-Evans relation as calibrated by \citet{nordgren02}}.
\end{figure}

\subsection{Interferometric observations}
The last few years have seen great advances in the area of interferometric observations of Galactic Cepheids. \citet{mourard97} measured the first angular diameter of a Cepheid ($\delta$~Cep) using the GI2T interferometer. Additional angular diameter measurements of other Cepheids were obtained by \citet{nordgren00} ($\alpha$~UMi, $\eta$~Aql, $\zeta$~Gem) with the NPOI interferometer and \citet{kervella01} ($\zeta$~Gem) with the IOTA interferometer. 

An important breakthrough was achieved by \citet{lane00}, who used the PTI interferometer to resolve for the first time the change in angular diameter of a Cepheid (in this case, $\eta$~Aql) as a function of pulsation period. \citet{kervella04c} extended the sample of Cepheids with such measurements to a total of seven objects, using the VLTI iterferometer. The combination of angular diameters and radial velocities as a function of phase enables the use of the interferometric Baade-Wesselink method \citep{sasselov94} to determine the distance to a Cepheid. \citet{kervella04a} used these distance determinations to calibrate the Cepheid Period-Radius and Period-Luminosity relations, although the sample size and period range is limited at the moment. However, future upgrades to the VLTI should result in 5\% distance determinations to $\sim 30$ Cepheids spanning a greater range in periods, enabling an accurate and precise independent calibration of the Cepheid Period-Luminosity relation.

\begin{figure}
{\centerline{\includegraphics[width=4.6in]{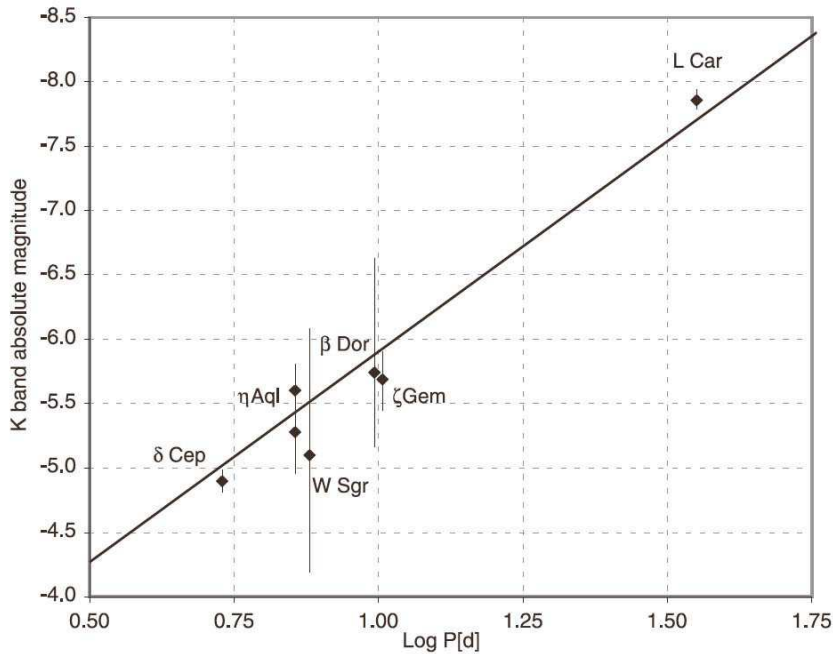}}}
\caption{Reproduction of Figure 2 from \citet{kervella04b}. K-band Period-Luminosity Relation for Galactic Cepheids with distances derived through interferometric observations.}
\end{figure}

\subsection{The Infrared Surface Brightness Technique}
\citet{wesselink69} derived a correlation between the surface brightness of a star in the $V$ band, $s_V$, and a dereddened color index, $(\bv)_0$, using stellar angular diameters measured through lunar occulations. The relation was applied to $\delta$~Cep to determine its absolute magnitude and hence, its distance. \citet{barnes76} used a larger sample of angular diameters to extend the method to later stellar types (more appropriate to Cepheids) and derived relations using the dereddened color indices $(\vr)_0$ and $(\ri)_0$. \citet{welch94} proposed a near-infrared variant of the technique, using the $K$-band surface brightness and the dereddened color index $(\vk)_0$.

\citet{fouque97} provided a modern recalibration of the surface brightness-color relations by using interferometrically determined angular diameters of giants and supergiants. The recent advances in interferometry discussed in \S2.2 have shown that Cepheids follow the same surface brightness-color relations as non-pulsating giants and supergiants \citep{lane02,kervella04b}. Indeed, \citet{kervella04d} were able to demonstrate that the angular diameter of $l$~Car as a function of phase follows the prediction of the \citet{fouque97} relation at the 1\% level.

\begin{figure}
{\centerline{\includegraphics[width=4.6in]{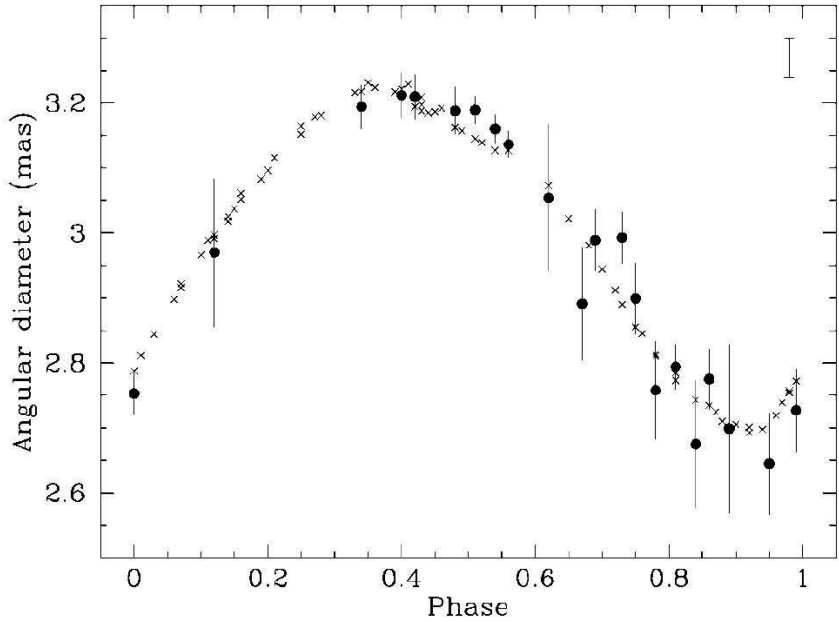}}}
\caption{Reproduction of Figure 3 from \citet{kervella04d}. Comparison of interferometrically-determined angular diameters for the Cepheid $l$~Car (filled circles) with angular diameters derived with the ISB technique (crosses) as calibrated by \citet{fouque97}.}
\end{figure}

\citet{gieren97} used the forementioned calibration of the surface bright\-ness-color relation to establish the use of the Infrared Surface Brightness (ISB) technique as a distance indicator. The ISB technique requires the adoption of a projection factor (hereafter, $p$-factor) to translate the measured radial velocities into pulsational velocities of the stellar surface. \citet{gieren97} determined distances to 16 Galactic Cepheids, and found excellent agreement with distances measured through the Zero-Age Main Sequence (ZAMS) method. \citet{gieren98} then applied the ISB technique to 34 Galactic Cepheids and derived Period-Luminosity relations in the $V$, $I$, $J$, $H$ and $K$ bands. The P-L relations exhibited statistically significant steeper slopes than the corresponding P-L relations of Cepheids in the Large Magellanic Cloud. 

Recently, \citet{tammann03} used over 300 Galactic Cepheids with $\bvi$ photometry from \citet{berdnikov00} to calculate Period-Color relations and compare them with model predictions. They also calculated Period-Luminosity relations using 25 Cepheids with ZAMS distances and 28 Cepheids with ISB distances, and confirmed the steeper slopes of the Galactic P-L relations (relative to LMC Cepheids). \citet{ngeow04} have also found similar results based on a slightly different set of Galactic Cepheids.

\begin{figure}
{\centerline{\includegraphics[width=4.6in]{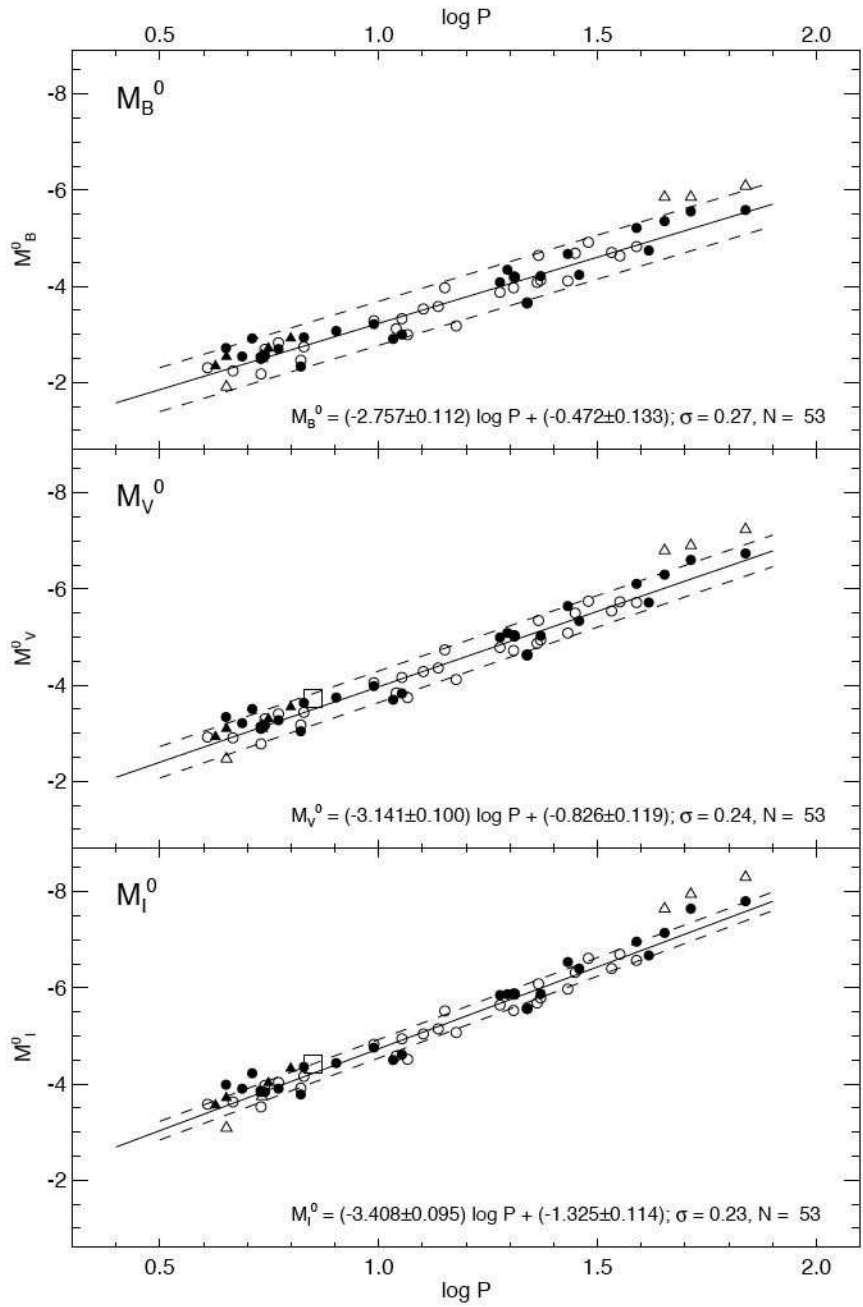}}}
\caption{Reproduction of Figure 11 from \citet{tammann03}. Period-Luminosity relations of Galactic Cepheid in $\bvi$ based on 25 objects with ZAMS distances (filled symbols) and 28 variables with ISB distances (open symbols). The square in the $V$ and $I$ panels represents the calibration derived by \citet{groenewegen00}.}
\end{figure}

This puzzling discrepancy in the slope of the P-L relations could have extremely severe consequences for the use of Cepheids as extragalactic indicators. However, as we shall see in \S3.1, recent work on LMC Cepheids seems to indicate that the adopted $p$-factor of the ISB technique may be to blame for the discrepancy. In this regard, the parallaxes being obtained by Benedict et al. can be combined with interferometric measurements to empirically determine the value and period dependence of the $p$-factor \citep{merand05}. Furthermore, theoretical modelling of Cepheid atmospheres \citep{nardetto04,marengo03,marengo02,bono00,sabbey95} will provide important insight into this problem.

\subsection{Other applications of Galactic Cepheids}
Galactic Cepheids can also be used to map the properties of our Milky Way. The detailed spectroscopic observations of \citet{andrievsky02a,andrievsky02b,andrievsky02c} and \citet{luck03} have been used to map the Galactic abundance gradient in great detail. This gradient can be compared with those obtained using older populations (such as planetary nebulae and open clusters) and younger populations (such as OB stars and \ion{H}{II} regions) to derive the rate of change of the gradient as a function of time \citep{maciel05}.

Galactic Cepheids also have, in many cases, extensive time series of photometric observations that go back as much as a century. Since the most massive Cepheids are expected to cross the instability strip (and hence evolve in period) in timescales of $\sim 10^3$ years, the existing synoptic data can be analyzed to measure the rate of change of the period of a Cepheid as a function of time. For example, \citet{turner04} have determined a period change for the 45-day Cepheid SV~Vul of $-214.3\pm5.5$ s yr$^{-1}$ and established that it is on its second crossing of the instability strip. As shown in that paper, the distribution of $\dot{P}$ as a function of period can be used in combination with semi-empirical and model predictions of period evolution to determine the particular crossing of the instability strip for any given Cepheid.

\begin{figure}[!hb]
{\centerline{\includegraphics[width=4.6in]{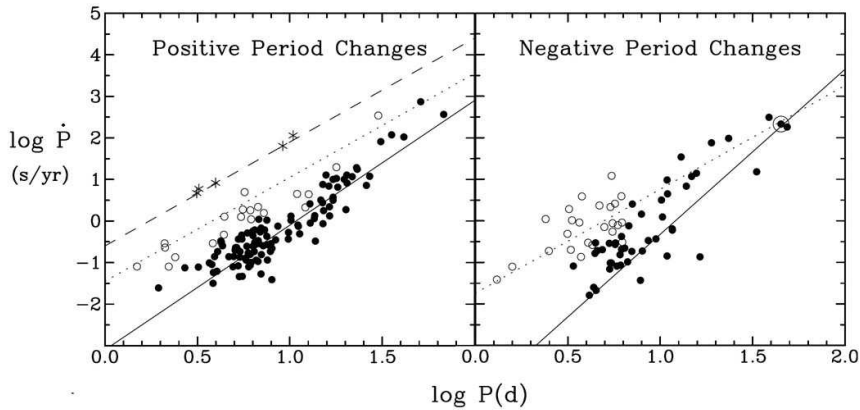}}}
\caption{Reproduction of Figure 2 from \citet{turner04}. Observed rates of period change for Cepheids and comparison with model and semi-empirical predicitions. Left: Cepheids with period increases, crossing the instability strip for the first (stars), third (filled) or fifth time (open). Right: Cepheids with period decreases, crossing the instability strip for the second (filled) or fourth time (open).}
\end{figure}


\section{The Magellanic Clouds}


\subsection{Cepheids in the Magellanic Clouds}

\subsubsection{Cepheid Period-Luminosity and Period-Color Relations}

\ \par

The history of Cepheids in the Magellanic Clouds extends a century into the past, since the discovery of the Cepheid Period-Luminosity relation by \citet{leavitt08} based on photographic observations of the Small Magellanic Cloud. Until a few years ago, the best $\bvrijhk$ Cepheid P-L relations of LMC Cepheids were those derived by \citet{madore91}, based on photoeletric and CCD observations of 25 ``Harvard Variables'' with $P>10$~d. \citet{sebo02} later augmented the classically observed $\bvri$ sample to $\sim 80$ Cepheids, while \citet{persson04} enlarged the $\jhk$ sample to $\sim 90$ variables.

The microlensing surveys (EROS, MACHO, OGLE) that began during the 1990s have obtained unprecedented synoptic data sets for the Magellanic Clouds, and have yielded literally {\it thousands} of new Cepheid variables (mostly at short periods). One of the first results to come out of these surveys was the determination by the MACHO team \citep{alcock95} of a clear separation between the P-L relations of fundamental and first-overtone pulsators.

\begin{figure}[!hb]
{\centerline{\includegraphics[width=4.6in]{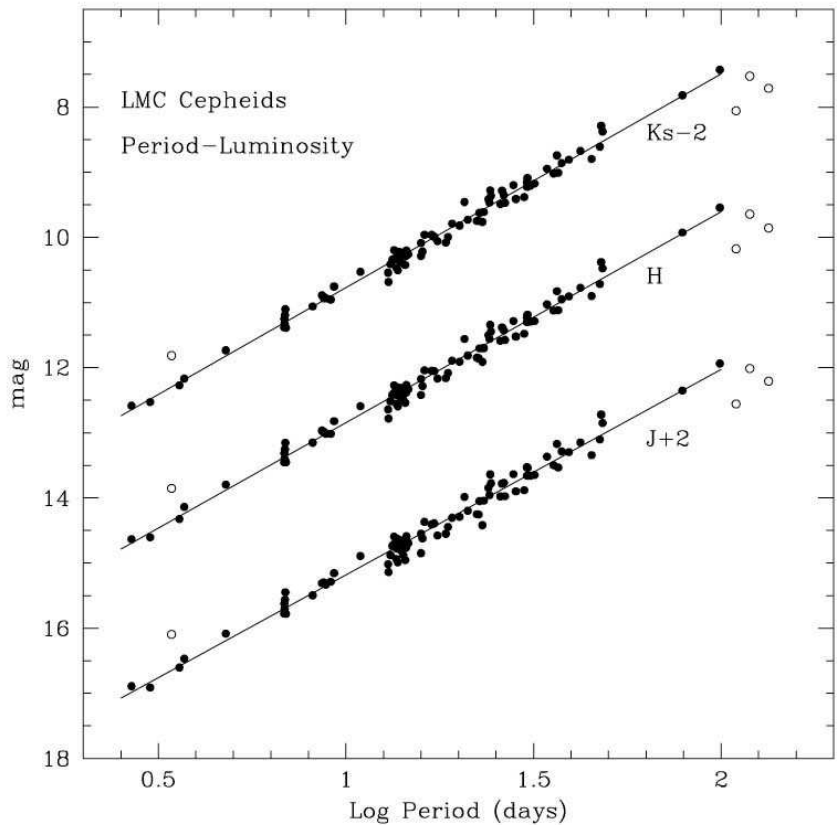}}}
\caption{Reproduction of Figure 3 from \citet{persson04}. Period-Luminosity-Color relations in $\jhk$ of Large Magellanic Cloud Cepheids}
\end{figure}

The OGLE team \citep{udalski99} derived new $\bvi$ P-L relations of Magellanic Cloud Cepheids using $>10^3$ variables discovered in their survey. Recenlty, \citet{sandage04} combined the observations of \citet{udalski99}, \citet{sebo02} and others to derive new Period-Luminosity and Period-Color relations of Magellanic Cloud Cepheids. They find evidence for a break in the slopes of these relations around $P=10$~d, a result that was confirmed by \citet{kanbur04}.

\begin{figure}[!hb]
{\centerline{\includegraphics[width=4.6in]{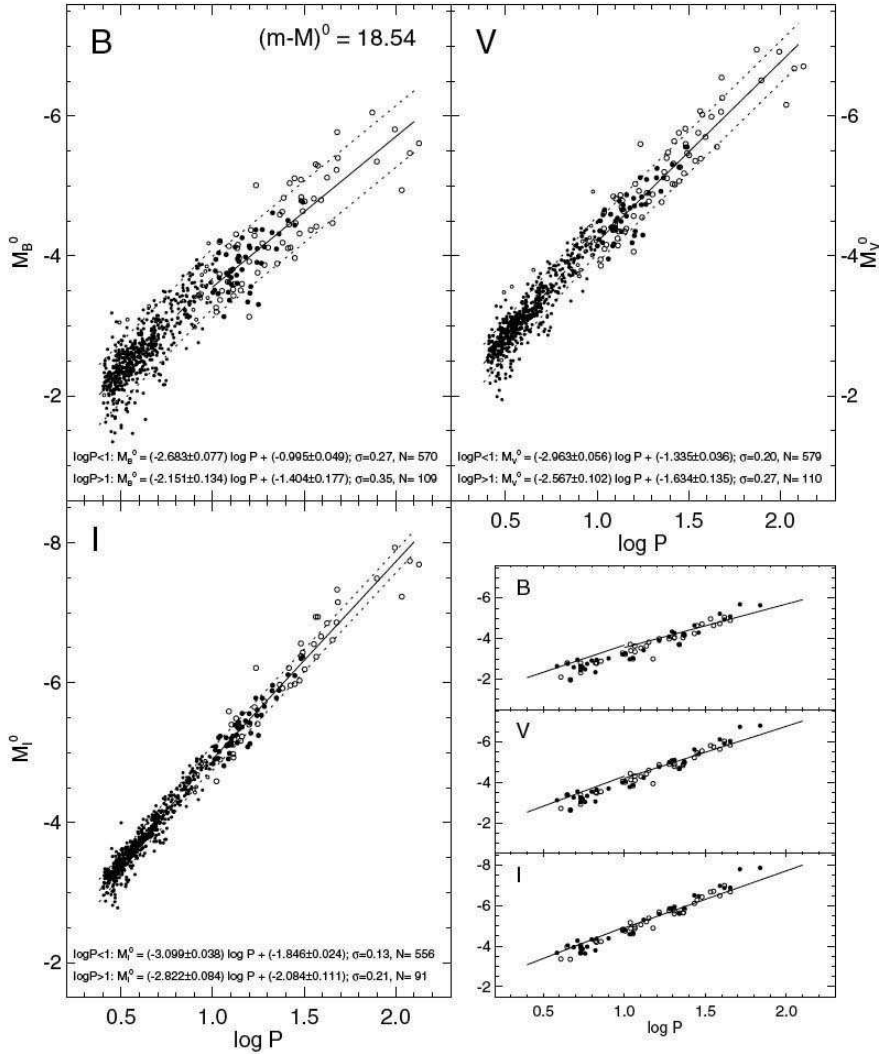}}}
\caption{Reproduction of Figure 4 from \citet{sandage04}. Period-Luminosity-Color relations in $\bvi$ of Large Magellanic Cloud Cepheids, fitted with two linear regressions breaking at $\log P=1.0$. For comparison, the lower right panel shows the Galactic Cepheids \citep[open and filled symbols, from][]{tammann03} overplotted on the mean LMC P-L relations.}
\end{figure}

\subsubsection{Determinations of the metallicity dependence of the Cepheid Distance Scale using Magellanic Cloud Cepheids}

\ \par

The application of the Cepheid Period-Luminosity Relation as a distance indicator crucially depends on the assumption of its universality from galaxy to galaxy. One possible systematic effect that could affect Cepheid distances at the 10\% level is the so-called ``metallicity effect'' --- that is, changes in the zeropoint and slope of the P-L and P-C relations as a function of chemical abundance. Several studies have taken advantage of the difference in mean chemical abundance between the Small and Large Magellanic Clouds ($\sim$ 0.5 dex) to characterize the effect using the Cepheids located in these galaxies.

\citet{sasselov97} used EROS photometry of SMC and LMC Cepheids to characterize the metallicity effect in the $B$ band at $\sim -0.4$~mag/dex. \citet{udalski01} relied on OGLE photometry of LMC, SMC and IC1613 Cepheids and other distance indicators (red clump stars, RR Lyrae variables, and the tip of the red giant branch) and found no discernible differential effects in the zeropoints of these methods. Recently, \citet{romaniello05} have obtained high-resolution spectroscopy of 37 Galactic and Magellanic Cloud Cepheids to investigate the metallicity effect. Their results seem to contradict the earlier empirical determination of \citet{kennicutt98} and favor the theoretical prediction of \citet{fiorentino02}.

\begin{figure}[!hb]
{\centerline{\includegraphics[width=4.6in]{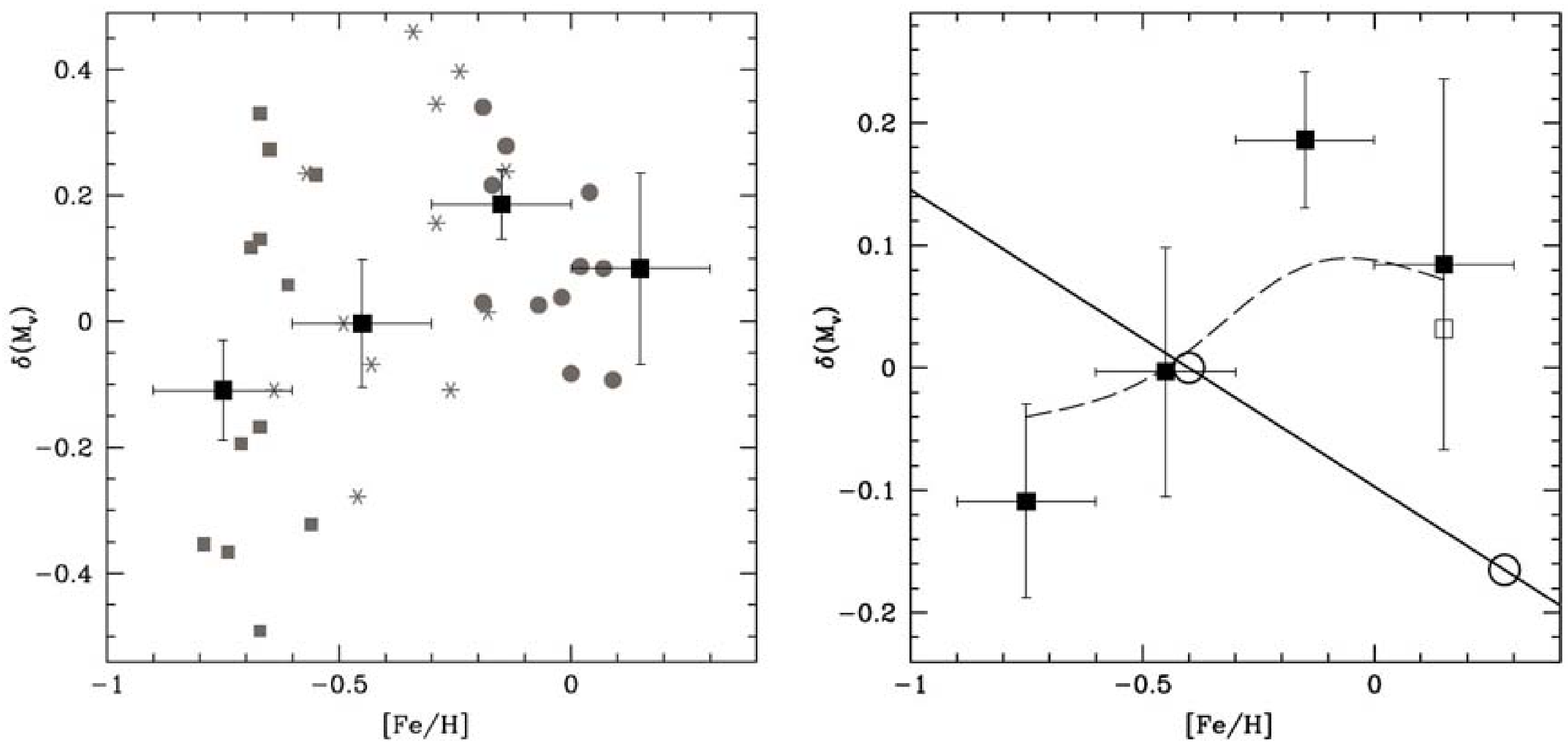}}}
\caption{Reproduction of Figure 1 from \citet{romaniello05}.Left: $V$-band residuals relative to the P-L relation adopted by \citet{freedman01}, plotted against [Fe/H], for Cepheids in the Galaxy (circles), the LMC (stars) and the SMC (squares). Median values are represented by black squares. Right: Comparison of the median values with the theoretical prediction of \citet{fiorentino02} (dashed line) and the empirical determination of \citet{kennicutt98} (open circles and solid line).}
\end{figure}

\subsubsection{Infrared Surface Brightness technique distances to Cepheids in the LMC}

\ \par

The latest development in Magellanic Cloud Cepheids comes from the application of the ISB technique to variables in the LMC. \citet{storm05} have obtained new near-infrared photometry and optical radial velocities for six Cepheids in the LMC cluster NGC$\,$1866, and have calculated ISB distances to these variables. \citet{gieren05} combined the \citet{persson04} light curves for seven Cepheids in the disk of the LMC with existing and new radial velocities to derive ISB distances to these objects.

The combined data set of ISB distances to LMC Cepheids, presented in \citet{gieren05}, shows a dependence of distance with period, although this correlation depends strongly on the assumption of a common distance for NGC$\,$1866 and the disk of the LMC \citep[see][for discussions on this point]{salaris03,storm05}. \citet{gieren05} ascribe the correlation of distances with periods to an erroneous choice of the $p$-factor, and calculate a new relation between $p$-factor and period with a steeper slope than what had been previously adopted. As a consequence of this revision, the slope of the Galactic $K$-band P-L relation is reduced and becomes fully consistent with the one derived by \citet{persson04} for LMC Cepheids.

While additional ISB distances to short-period Cepheids in the disk of the LMC are required to strengthen the new $p$-factor relation proposed by \citet{gieren05}, it is comforting to see that its adoption clears the puzzling discrepancy in P-L slopes mentioned at the end of \S2.3.

\begin{figure}[!hb]
{\centerline{\includegraphics[width=4.6in]{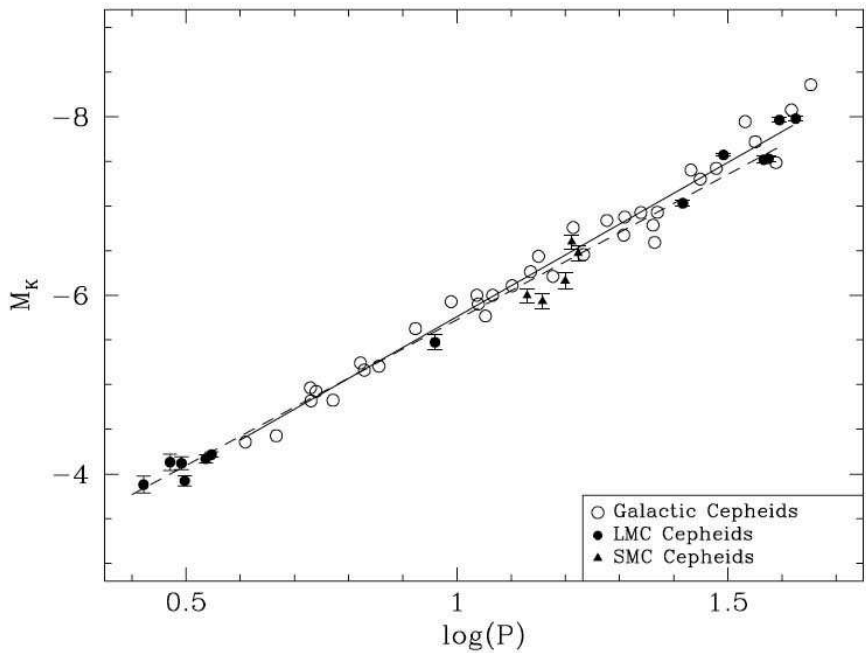}}}
\caption{Reproduction of Figure 12 from \citet{gieren05}. The application of the new $p$-factor scaling has reconciled the slopes of the P-L relations in the Galaxy and the Magellanic Clouds.}
\end{figure}

\vfill\pagebreak

\subsection{Eclipsing binaries in the Magellanic Clouds}

Our detailed knowledge of Magellanic Cloud Cepheids reviewed in the previous pages was, until recently, hampered by one crucial piece of information that was relatively unknown: accurate and precise distances to the Magellanic Clouds. The review by \citet{gibson00} presented a disturbing picture of the situation at the beginning of the decade, with a spread of $\sim$ 0.5 mag among different estimates of the distance modulus of the LMC and little overlap of error bars --- tell-tale signs of neglected systematics.

One of the most promising solutions to this problem is the determination of geometrical distances to the Clouds, based on observations of detached eclipsing binaries (hereafter, DEBs). While this method requires the use of theoretical stellar atmosphere models (which could contribute unknown systematic errors), the risk can be mitigated by studying many systems with a large range of effective temperatures.The line-of-sight extinction, another source of possible systematic error, can be accurately characterized by obtaining UV spectroscopy of the systems.

The first-generation microlensing surveys (EROS, MACHO, OGLE-I) discovered new promising early-type EBs and provided high-quality light curves for previously-known systems. Thus far, three DEBs in the LMC have been studied in detail: HV$\,$2274 \citep{guinan98}, HV$\,$982 \citep{fitzpatrick02} and EROS$\,$1044 \citep{ribas02}. They yield very consistent distances to the center of the LMC of 48$\pm$2~kpc. A fourth system, HV$\,$5936 \citep{fitzpatrick03}, consists of a semi-detached eclipsing binary that appears to lie a few kpc on the foreground of the LMC disk. Two more systems are under analysis at the present time: EROS 1066 and MACHO053648.7-691700.

\begin{figure}[!hb]
{\centerline{\includegraphics[width=4.6in]{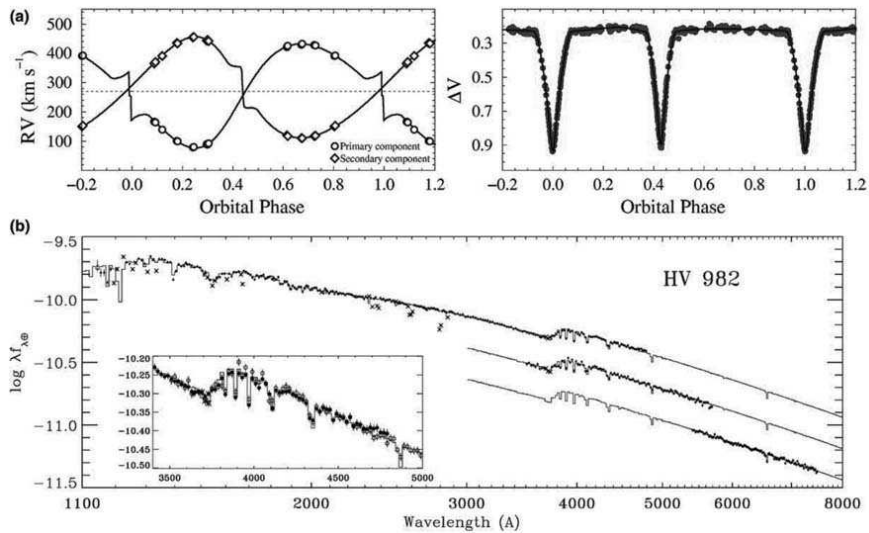}}}
\caption{Reproduction of Figure 4 from \citet{guinan04}, showing the radial velocity curve, $V$-band light curve, and spectrophotometric data for the LMC DEB HV$\,$982.}
\end{figure}

\citet{harries03} recently presented first results from an ongoing program to study over 100 eclipsing binaries in the SMC, using OGLE photometry and radial velocities obtained in multiplexed fashion with the 2dF spectrograph. They derived a distance of 60$\pm$4~kpc to the SMC, primarily based on 3 DEBs and 6 semi-detached systems. \citet{wyithe02} had previously advocated the use of semi-detached systems over DEBs for distance determination because the increased ease of analysis \citep[see also][for additional details]{wilson04}.

Another interesting and recent development was the study by \citet{lepischak04} of a binary system in the LMC discovered by the MACHO survey \citep{alcock02} in which the primary star is a 2-day overtone Cepheid and the secondary star appears to be a late-type giant. Further observations in the near-infrared are required before this promising system can be fully caracterized. 

\section{The Local Group}

\subsection{M31 and M33}

M31 and M33 have been surveyed for variables on many occasions, starting with the seminal work of \citet{hubble26,hubble29} and \citet{baade63,baade65}. In one of the first CCD studies of extragalactic Cepheids, \citet{freedman90,freedman91} obtained BVRI photometry of variables discovered by the earlier photographic surveys and showed the application of multi-wavelength observations to account for internal extinction in distance determination.

The first CCD-based synoptic survey for Cepheids and eclipsing binaries in M31 and M33 was carried out by the DIRECT project. The first results for M31 were presented in \citet{kaluzny98,stanek98}, while \citet{macri01a} did the same for M33. In all the DIRECT project surveyed $\sim 0.5\sq \deg$ of M31 and $\sim 0.3 \sq \deg$ of M33 over four years and discovered $\sim 10^3$ Cepheids, $\sim 200$ eclipsing binaries, and hundreds of miscellaneous variables \citep[see][for the latest results in M31]{bonanos03a}. As an interesting side note, the brightest and longest period ($P\sim 55$~d) Cepheid that had been discovered by \citet{hubble26} in M33 was found to have stopped pulsating at the present time and increased in brightness by $\sim 0.5$~mag at $B$ \citep{macri01b}.

Several promising DEBs were discovered in M31 and M33. Follow-up photometry of was obtained with the Kitt Peak 2.1-m telescope, while spectroscopic measurements were carried out using the Keck and Gemini North telescopes. \citet{bonanos05} recently presented the analysis of the first DEB in M33, which resulted in a determination of the distance to that galaxy with a precision of 8\%. Meanwhile, \citet{ribas04} are carrying out a similar follow-up program of DIRECT DEBs in M31, and \citet{todd05} have announced the discovery of additional eclipsing binaries in M31.

In another follow-up study to the DIRECT survey, \citet{macri05a} have obtained single-phase $\jhk$ photometry for $\sim 200$ Cepheids in M33. The data will combined with additional optical observations to attempt a differential characterization of the metallicity dependence of the Cepheid P-L relation. This is possible thanks to the large abundance gradient of M33, which spans $\sim 0.5$~dex.
\pagebreak

\begin{figure}[!ht]
{\centerline{\includegraphics[width=4.6in]{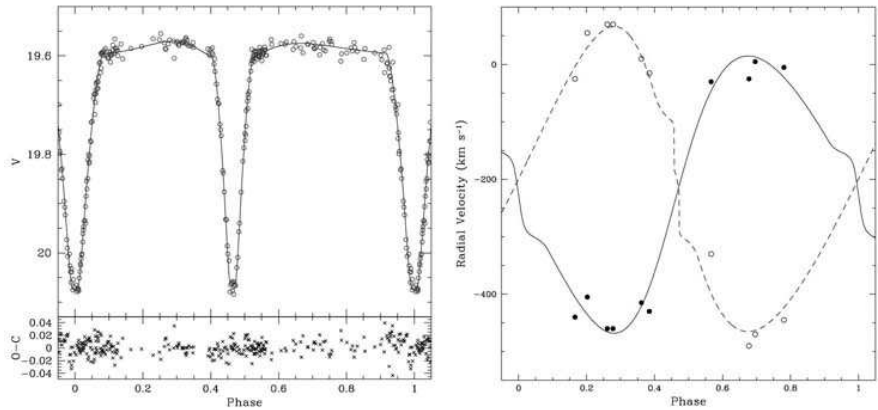}}}
\caption{Reproduction of Figures 2 and 4 from \citet{bonanos05}, showing the $V$-band light curve and radial velocity curve of a DIRECT eclipsing binary in M33.}
\end{figure}

\begin{figure}[!hb]
{\centerline{\includegraphics[width=4.6in]{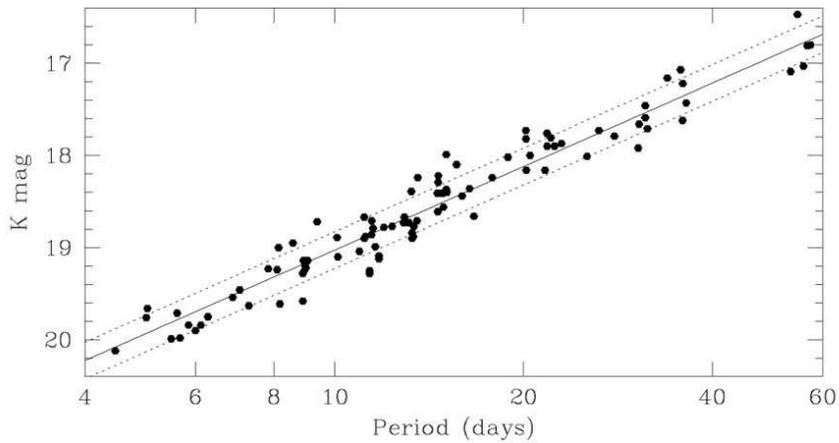}}}
\caption{Preliminary $K$-band P-L relations of Cepheids in M33, from \citet{macri05a}.}
\end{figure}

The Infrared Surface Brightness technique could -- in principle -- be applied to Cepheids in M31 and M33. \citet{forestell04} have presented preliminary results from spectroscopic observations of Cepheids in these galaxies using the Hobby-Ebberly Telescope. Near-IR light curves do not exist for these Cepheids yet, but could be obtained with a modest investment of observing time at 8-10m class telescopes.

\vfill\pagebreak

\subsection{Other Local Group galaxies}

CCD surveys for Cepheids in other Local Group galaxies have also been carried out, most notably in the case of IC 1613 \citep{antonello02a,udalski01} and NGC$\,$6822 \citep{antonello02b,pietrzynski04a,baldacci05}.

A few dozen eclipsing binaries have also been discovered in other Local Group galaxies, such as Fornax, Leo I, NGC$\,$6822 and Carina \citep[for details, see the recent review by][]{guinan04}. Their number are likely to grow in the coming years as additional synoptic surveys are carried out.

\section{Beyond the Local Group}

While the main focus of this review is on Local Group Cepheids and EBs, I would like to present a short summary of recent work done beyond 1~Mpc. The ARAUCARIA project \citep{pietrzynski04b} is carrying out a comprehensive ground-based survey of southern hemisphere galaxies, including several members of the Sculptor Group (located at $\sim$2~Mpc) with the aim of improving the calibration of several distance indicators. As part of their survey, they have recently discovered over one hundred Cepheids in the Sculptor Group galaxy NGC$\,$300 \citep{gieren04}.

The availability of 8-10m class telescopes and the possibility of obtaining data in ``service'' or ``queue'' mode has enabled ground-based searches for Cepheids at even greater distances. Recently, \citet{thim03} reported the discovery of Cepheids in M83, at a distance of $\sim$ 4.5~Mpc using the VLT. A reanalysis of that data by \citet{bonanos03b} demonstrated the power of difference-image analysis \citep[][hereafter DIA]{alard00} for increasing the number of detected Cepheids. Unfortunately, DIA only provides the ``AC term'' of variable lightcurves, so \citet{bonanos03b} proposed a ``hybrid'' approach for future Cepheid work, combining difference-image analysis of ground-based synoptic surveys with follow-up imaging from HST (to obtain the absolute flux level of the variables).

Moving even further from the Local Group, we come to a unique galaxy in terms of its importance for the Extragalactic Distance Scale. The distance to the spiral galaxy NGC$\,$4258 can be determined to very high precision and accuracy ($\sim 3\%$) through geometrical methods based on VLBI observations of water masers located in the accretion disk of its central supermassive black hole \citep{herrnstein99,humphreys05}. Thus, it can provide an excellent determination of the zeropoint of the Cepheid P-L relation. A survey for Cepheids in this galaxy was carried out by \citet{maoz99} using the WFPC2 camera onboard {\it HST} \citep[see][for a revision to the original result]{newman01}. 

Unfortunately, the size and period range of the HST/WFPC2 Cepheid sample in NGC$\,$4258 was rather limited. For this reason, an additional survey was recently carried out using the newer Advanced Camera for Surveys on {\it HST} \citep{macri05b}. Two fields were targeted: an ``outer'' one with LMC-like chemical abundance, and an ``inner'' one with an abundance close to solar. The two samples should allow another differential measurement of the metallicity effect of the Cepheid P-L relation, similar to the one carried out in M101 by \citet{kennicutt98} using HST/WFPC2.

The larger field of view of ACS, increased quantum efficiency and better spatial sampling helped to increase the size of the Cepheid sample by a factor of 10 relative to that of \citet{maoz99}. Follow-up near-IR observations of a sub-sample of these variables using NICMOS on {\it HST} are under way. Lastly, a ground-based survey using the Gemini North 8-m telescope has recently been completed by myself and collaborators, with the aim of increasing the number of long-period ($P>40$~d) Cepheids in this galaxy, which will be crucial to obtain an independent measurement of the slope of the P-L relation and test its universality, as discussed previously in the text.

\begin{figure}[!hb]
{\centerline{\includegraphics[width=4.6in]{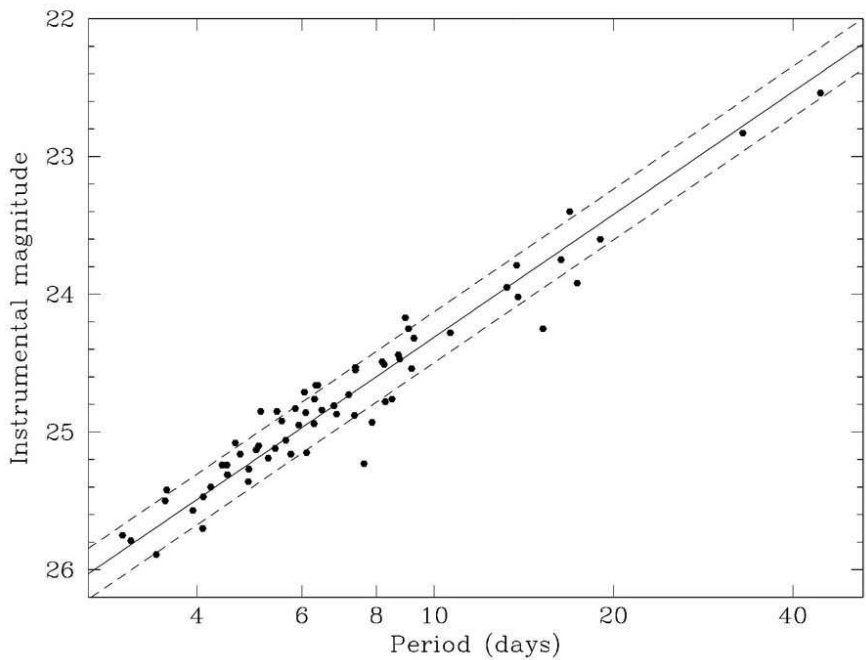}}}
\caption{Preliminary $I$-band P-L relations of Cepheids discovered in an outer field of NGC$\,$4258 using HST/ACS, from \citet{macri05b}.}
\end{figure}

\section{Conclusions}

I hope this review has shown the great diversity and large number of ongoing observational work on Cepheids and eclipsing binaries in our Local Group (and beyond). These investigations should yield -- in the near future -- a more robust absolute calibration of the Extragalactic Distance Scale and a clearer understanding of its systematic uncertainties. Such an improvement will help reduce the error budget of the determination of the Hubble constant $H_0$ \citep{freedman01} and increase our understanding of fundamental stellar physics.



\bibliography{./macri}
\end{document}